\documentclass[aps,prl,showpacs,twocolumn]{revtex4-1} 	%looks like final product
\usepackage{graphicx, amsmath, amssymb}
\usepackage{epstopdf}		%convert eps figures to pdf for pdflatex, if needed
\newcommand{\super}[1]{$^{\text{#1}}$}	% superscript in text, with math font sizing
\newcommand{\sub}[1]{$_{\text{#1}}$}	% subscript in text, with math font sizing
\newcommand{\degree}{\ensuremath{^\circ}}	% degrees notation for temperature

\begin{document}

%% == Definitions ==
\title{Hyperfine frequencies of $^{87}$Rb and $^{133}$Cs atoms in Xe gas}
\author{B. H. McGuyer}%
\author{T. Xia}
\altaffiliation{Present address:  Department of Physics, University of Wisconsin, Madison, Wisconsin 53706, USA.}
\author{Y.-Y. Jau}
\altaffiliation{Present address:  Sandia National Laboratories, Albuquerque, New Mexico 87123, USA.}
\author{W. Happer}
\affiliation{Department of Physics, Princeton University, Princeton, New Jersey 08544, USA}
\date{\today}%

\begin{abstract}
The microwave resonant frequencies of ground-state $^{87}$Rb and $^{133}$Cs atoms in Xe buffer gas are shown to have a relatively large nonlinear dependence on the Xe pressure, presumably because of RbXe or CsXe van der Waals molecules. The nonlinear shifts for Xe are opposite in sign to the previously measured shifts for Ar and Kr, even though all three gases have negative linear shifts. The Xe data show striking discrepancies with the previous theory for nonlinear shifts. Most of this discrepancy is eliminated by accounting for the spin-rotation interaction, $\gamma{\bf N}\cdot {\bf S}$, in addition to the hyperfine-shift interaction, $\delta\! A\,{\bf I}\cdot{\bf S}$, in the molecules.  To the limit of our experimental accuracy, the shifts of $^{87}$Rb and $^{133}$Cs in He, Ne, and N$_2$ were linear with pressure.
\end{abstract}

%Suggested PACS numbers
% 32.70.Jz	Line shapes, widths, shifts
% 32.30.Bv	Radio-frequency, microwave, and infrared spectra
% 32.30.Dx	Magnetic resonance spectra
% 34.20.Cf	Interatomic potentials and forces
% Microwave optical double resonance spectroscopy, 33.40.+f

\maketitle

%%%%%
 Frequency standards based on the microwave resonant frequencies of optically pumped alkali-metal atoms in buffer gases are still the most widely used atomic clocks. Applications include satellite navigation systems, base stations for mobile phone networks, and laboratory frequency synthesizers \cite{camparo:2007pt}. These are secondary frequency standards because the resonant frequencies $\nu=\nu(T,p_1,p_2,\ldots)$ are not the same as the free-atom frequencies $\nu_0$, but depend on the partial pressures $p_1, p_2,\ldots$ of the buffer gas  (e.g., mixtures of Ar and N$_2$)  and on the  temperature $T$.   Chemically inert buffer gases are necessary to keep the optically pumped clock atoms from diffusing too quickly to the cell walls. This would shorten the coherence time of the ``0--0" resonant frequency, broaden the resonance line, and seriously degrade the clock performance \cite{vanier:1989}. The frequency shifts, $\nu-\nu_0$, can be measured very precisely, and
 are one of the few measurable phenomena that can provide information about spin interactions in loosely-bound van der Waals (vdW) molecules, which are of considerable current interest \cite{walker:1997, Brahms:2010}. Here we present surprising new results from precision measurements of the shifts of $^{87}$Rb and $^{133}$Cs atoms in  Xe gas.

Gong {\it et al.}\ \cite{gong:2008} showed that one would expect the pressure shifts from Xe to have a nonlinear dependence on the Xe pressure $p$ because of the formation of vdW molecules in three-body collisions between two Xe atoms and a Rb atom.  For a Rb atom bound in a RbXe molecule the precession of  the nuclear spin ${\bf I}$ and the electron
spin ${\bf S}$ about each other is perturbed  by the hyperfine-shift interaction
\begin{align}	\label{hfs}
V_\text{hfs} & = \delta A \: \mathbf{I}\cdot \mathbf{S}.
\end{align}
The interaction (\ref{hfs}) acts until the molecule is broken up in a  collision with another Xe atom. The same interaction (\ref{hfs}) acts impulsively when an alkali-metal atom has a binary collision with a buffer-gas atom or molecule.  Following Gong {\it et al.}\ \cite{gong:2008}, we fit our measured frequencies $\nu$ to a theoretical function of pressure,
\begin{align}	\label{f0}
f_0=\nu_0+sp+\Delta^2\nu,
\end{align}
where the nonlinear shift is
\begin{align}	\label{FeiModel}
\Delta^2\nu &= - \left( \frac{1}{2 \pi T} \right) \frac{\phi^3}{1 + \phi^2}.
\end{align}
The theoretical function (\ref{f0}) is characterized by four parameters: (i) the free-atom frequency $\nu_0$, which varies daily by a few Hz because of drifts in the ambient magnetic field of our laboratory; (ii) the ``slope" $s$, which represents the linear shifts due to binary collisions and three-body collisions at high pressures \cite{gong:2008}; (iii) the formation rate $1/T\propto p^2$ of vdW molecules in three-body collisions; and (iv) the differential phase shift $\phi = \delta A[I]\tau/(2\hbar)\propto 1/p$ of the atomic coherence after a mean bound-atom lifetime $\tau$, where $[I] = 2I+1$ denotes the number of sublevels for a spin of quantum number $I$.

%	FIGURE 1
\begin{figure}[t!]
	\centering
	\includegraphics[width=8.5cm]{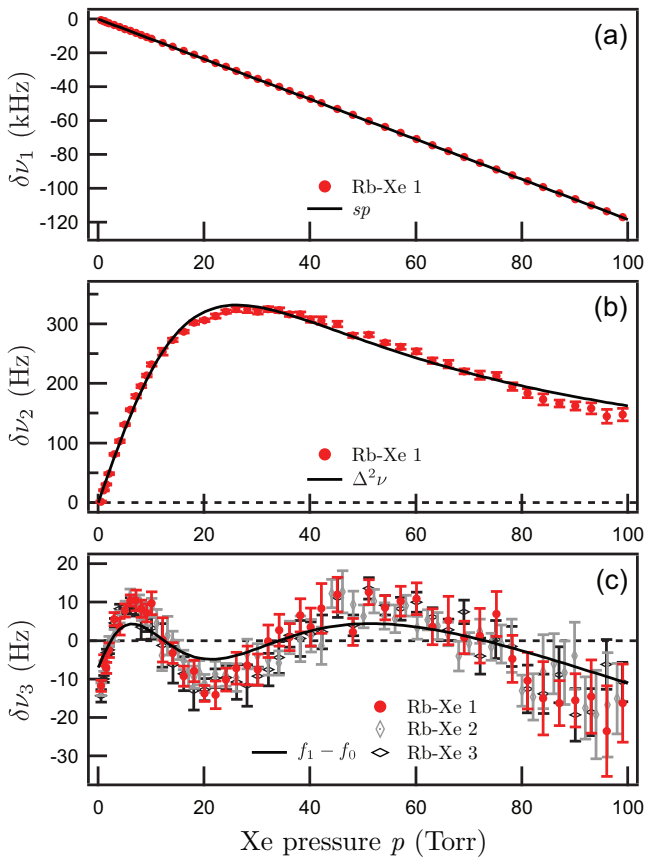}
	\caption{\label{fig:1}(Color online) Measured ``0--0" resonant frequencies $\nu$ of \super{87}Rb in Xe at 40.0\degree C and $B=1$ G.
	(a) The shift $\delta\nu_1 = \nu - \nu_0$ from the free-atom frequency $\nu_0\approx6.834683$~GHz.  The solid line is the linear, limiting shift $sp$ at high pressures.
	(b) The nonlinear shift $\delta \nu_2 = \delta\nu_1 -sp$.  The solid curve is the nonlinear shift $\Delta^2\nu$ of (\ref{FeiModel}).
	(c) The fit residuals to (\ref{f0}),  $\delta\nu_3 = \delta\nu_2-\Delta^2\nu=\nu-f_0$.
The solid curve is the difference, $f_1-f_0$ of the best-fit theoretical curves, with and without the spin-rotation interaction (\ref{sr}).
Three separate data sets demonstrate the repeatability of the measurements.
	}
\end{figure}

In Fig.~1(a) we plot the measured shift $\delta\nu_1=\nu-\nu_0$ of $^{87}$Rb in Xe at $40.0$\degree C. The free-atom frequency $\nu_0$ was obtained by fitting the data to the model (\ref{f0}). The solid curve is $sp$, the linear part of (\ref{f0}). The apparent linearity of the shift $\delta\nu_1$ is deceptive, as shown in Fig.~1(b), where we plot $\delta \nu_2=\delta\nu_1-sp=\nu-\nu_0-sp$. This reveals the highly nonlinear nature of the shift. The solid curve in Fig.~1(b) is the nonlinear shift $\Delta^2\nu$ of (\ref{f0}). Once more, the apparently good agreement between the measured $\delta\nu_2$ and the predicted $\Delta^2\nu$ is deceptive, as shown in Fig.~1(c), where we plot the residuals $\delta\nu_3=\delta\nu_2-\Delta^2\nu=\nu-f_0$. The residuals $\delta\nu_3$ have a striking, oscillatory behavior, which is the same for three independent sets of measurements.
Similar residuals $\delta \nu_3$ are seen with $^{133}$Cs in Xe.

To derive the nonlinear shift (\ref{FeiModel}), Gong {\it et al.}\ \cite{gong:2008} assumed that vdW molecules form in a single vibration-rotation state in which the evolution of ${\bf I}$ and ${\bf S}$ is perturbed by the hyperfine-shift interaction (\ref{hfs}).
Two problems with this assumption are: (i) the vdW molecules can form in many (hundreds) of vibration-rotation states \cite{pritchard:1976}, each with different values of the shift parameter $\delta A$ and of $T$ and $\phi$; and (ii) the effects of spin interactions other than (\ref{hfs}), such as the spin-rotation interaction 
\begin{align}	\label{sr}
V_\text{sr} & = \gamma \, \mathbf{N}\cdot \mathbf{S},
\end{align}
between the rotational angular momentum $\mathbf{N}$ of the molecule and $\mathbf{S}$, are ignored.  
The interaction (\ref{sr}) is particularly large for RbXe or CsXe molecules, and is equivalent to a magnetic field, of roughly 38~G for RbXe, oriented along {\bf N} \cite{wu:1985, bouchiat:1972}. 
Both interactions (\ref{hfs}) and (\ref{sr}) need to be included in an improved theory of nonlinear pressure shifts.  

Here, we show that for $^{87}$Rb or $^{133}$Cs in Xe, most of the discrepancy between theory and experiment is eliminated by including the effects of the spin-rotation interaction (\ref{sr}) in a model that still has only one vibration-rotation state. The function (\ref{f0}) must be replaced by
\begin{align}	\label{f1}
f_1=\nu_0+sp+\Delta^2_{1}\nu,
\end{align}
where the nonlinear shift that includes the effects of both interactions (\ref{hfs}) and (\ref{sr}) is
\begin{align}	\label{LowFieldModel}
\Delta_{1}^{2}\nu &= - \left(\frac{1}{2 \pi T} \right)\sum_{\sigma=-2I}^{2I} \frac{W_\sigma (1+r_1 \sigma)^3 \phi^3}{1 +  (1+r_1 \sigma)^2 \phi^2}.
\end{align}
The parameter $r_1=2\gamma N/(\delta A[I]^2)$ accounts for the spin-rotation interaction (\ref{sr}).
We will discuss the meaning of the remaining symbols of (\ref{LowFieldModel}) and its derivation below.

We refit our data to the revised function (\ref{f1}), where we constrained the new parameter $r_1$ by estimating $r_1 \phi p = \gamma N \tau p / ([I] \hbar)$ from measurements of $\langle \gamma N \rangle$ and $\langle \tau p \rangle$ for RbXe by Bouchiat {\it et al.}\ \cite{bouchiat:1972}.
We do not show the fits of (\ref{f1})  corresponding to Fig.~1(a)--(b), since they look very similar to the fits of (\ref{f0}).
However, as Table \ref{tab:1} shows, the fits to (\ref{f1}) give substantially different values for the parameters  $T$ and $\phi$.
In Fig.~1(c) the solid line is the difference, $f_1-f_0$ between the two theoretical fit functions. The revised curve $f_1$ gives substantially smaller residuals, and displays the same oscillatory behavior with pressure.  
Similar results are seen with $^{133}$Cs in Xe.  
The remaining residuals are probably due to the distribution of vibration-rotation states of the vdW molecules, or the neglect of still smaller spin interactions, such as the anisotropic hyperfine-shift interaction or the electric quadrupole interaction.  

%	TABLE 1
\begin{table}[t!]%[H] add [H] placement to break table across pages
\caption{\label{tab:1}Fit parameters for the pressure shifts of \super{87}Rb at 40\degree C and $B = 1$ G and of \super{133}Cs at 35\degree C and $B = 0.2$ G.
Uncertainties for the slopes $s$ are typically $\pm0.25\%$.}
\begin{ruledtabular}
\begin{tabular}{l c c c c c}
Metal & Gas 			& $r_1 \phi p$ 	& $Tp^2$  			& $\phi p$ 		& $s$  \\
			&		& rad$\:$Torr	& sec$\:$Torr$^2$		& rad$\:$Torr		& Hz$\:$Torr$^{-1}$\\
\hline
\super{87}Rb 	& Xe 		& 7.97		& $0.082 \pm 0.010$	& $-14.8 \pm 1.5$	& $-1184.0$\\
			& Xe		& --			& $0.164 \pm 0.013$	& $-26.2 \pm 1.2$	& $-1183.7$\\
			& Ar 		& -- 			& $0.070 \pm 0.012$ 	& $2.21 \pm 0.16$	& $-53.71$\\
			& Kr 		& -- 			& $1.08 \pm 0.31$ 		& $12.5 \pm 1.9$	& $-558.1$\\
			& He			&&&&	$714.2$\\
			& Ne			&&&&	$387.3$\\
			& N\sub{2}	&&&&	$518.0$\\
\super{133}Cs & Xe   	& $2.79$ 		& $0.0185$			& $-5.96$ 			& $-2243.9$\\
			& Xe		& --			& $0.059 \pm 0.006$	& $-15.8 \pm 1.3$ 	& $-2242.4$\\
			& He			&&&&	$1141.9$\\
			& Ne			&&&&	$579.9$\\
			& N\sub{2}	&&&&	$828.6$
\end{tabular}	
\end{ruledtabular}
\end{table}

To understand the origin of the formulas (\ref{FeiModel}) and (\ref{LowFieldModel}), note that when a clock atom is captured in a vdW molecule, the microwave coherence frequency changes slightly.  Gong {\it et al.}\ \cite{gong:2008} ignored the spin-rotation interaction (\ref{sr})  and assumed that the coherence frequency was shifted by the amount $\delta\omega=\delta A[I]/(2\hbar)$, leading to the mean phase shift $\phi = \delta \omega \, \tau$.   The form of the function (\ref{FeiModel}) describes how a statistical ensemble of molecular lifetimes shifts the resonant frequency $\nu$.

The generalization (\ref{LowFieldModel}) is a superposition of functions with the same form as  (\ref{FeiModel}), each labeled by an integer $\sigma$, with a weight $W_{\sigma}$ and with a mean phase shift $\phi\to (1+\sigma r_1)\phi$.
Assuming the interaction $g_S\mu_B S_z B$ ($g_S$ is the electronic $g$ factor and $\mu_B$ is the Bohr magneton) with an external magnetic field $B$ is negligible compared to the spin-rotation interaction (\ref{sr}), the ground-state energy sublevels $|fm\rangle$ for a bound clock atom will be quantized with the azimuthal quantum number $m$ along the rotational angular momentum ${\bf N}$.
The total spin angular momentum quantum number is  $f=a=I+1/2$ or $f=b=I-1/2$. The interactions (\ref{hfs}) and (\ref{sr}) shift the sublevel energies as shown in Fig.~2. We can write the shift in the Bohr frequency of a coherence between sublevels $|a m\rangle$ and $|b,\sigma-m\rangle$ as $\delta\omega_{\sigma}=(\delta E_{a m}-\delta E_{b,\sigma-m})/\hbar = \delta A[I](1+r_1\sigma)/(2\hbar)$.
The weights account for the fraction of the unbound-atom coherence that evolves with the frequency shift $\delta\omega_{\sigma}$ in the molecule.  Averaging over an isotropic distribution of quantization directions gives the weights as the sum of Clebsch-Gordan coefficients, % \cite{Varshalovich},
\begin{align}
W_\sigma =  \sum_{\mu k} \frac{(-1)^\sigma}{[k]}C_{a0;a0}^{k0} C_{a,\mu;a,-\mu}^{k0} C_{b0;b0}^{k0} C_{b,\sigma-\mu; b, \mu-\sigma}^{k0}.
\label{Ws}
\end{align}
For $^{87}$Rb with $I=3/2$, (\ref{Ws}) becomes { $[W_3,\ldots,W_{-3}] = [9, 9, 23, 23, 23, 9, 9]/105$
\footnote{For \super{133}Cs, $[W_7,\ldots,W_{-7}]$ = [245, 245, 425, 425, 659, 659, 1231, 1231, 1231, 659, 659, 425, 425, 245, 245]/9009.}.
The weights depend on the choice of the unbound-atom hyperfine coherence, so the nonlinear shifts (\ref{LowFieldModel}) will be different for clocks that do not use the 0--0 transition, for example, ``end-resonance'' clocks \cite{jau:2004} or lin$\parallel$lin clocks \cite{linparlin:2005}.
The weights satisfy $\sum_{\sigma}W_{\sigma}=1$, so if we let $r_1\to 0$ in (\ref{LowFieldModel}) we recover (\ref{FeiModel}).  We also recover (\ref{FeiModel}) in the limit that the spin-rotation interaction (\ref{sr}) is negligible compared to the interaction $g_S\mu_B S_z B$.
Between the small-$B$ and large-$B$ limits, the nonlinear shifts from vdW molecules will depend on the applied field $B$.  In all cases, the analysis leading to (\ref{LowFieldModel}) predicts a molecular contribution $s_\text{m} = \phi /(2 \pi T p)$ to the slope $s$ of (\ref{f0}) or (\ref{f1}).

%	FIGURE 2
\begin{figure}[t]
	\centering
	\includegraphics[width=8.5cm]{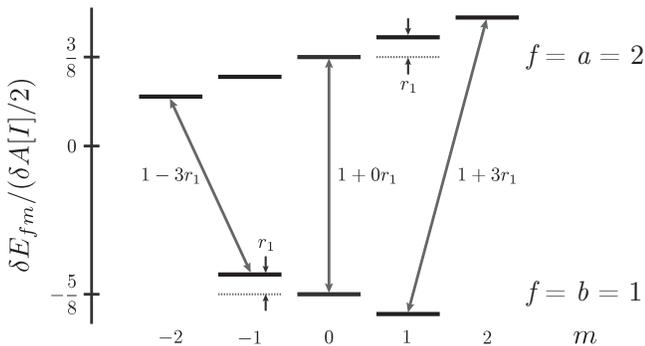}
	\caption{\label{fig:2} Energy shifts of the sublevels of a $^{87}$Rb atom in a vdW molecule due to the interactions (\ref{hfs}) and (\ref{sr}). Representative  perturbations $1+r_1\sigma$ of the coherence frequencies are shown. The sublevels are quantized along ${\bf N}$.  }
\end{figure}

We measured the 0--0 resonant frequencies $\nu$ of $^{87}$Rb and $^{133}$Cs in pure buffer gases with two laser-pumped, vapor-cell clock systems that closely follow the design of Gong {\it et al.}\ \cite{gong:2008}.  The systems are based on two feedback loops:  one to lock the carrier frequency of frequency-modulated microwaves to the 0--0 transition, and another to lock the optical frequency of pumping light to produce no light shift of the 0--0 transition.
A frequency counter with 1~Hz precision, referenced to a Rb frequency standard, sampled the locked carrier to provide $\nu$.
We estimated the uncertainty with the sample standard deviation, which increased with the 0--0 linewidth at higher pressures.
For the Cs system, the implementation of microwave frequency modulation limited the Xe pressure range and led to increased carrier noise.

The interactions (\ref{hfs}) and (\ref{sr}) are so large for RbXe molecules that we needed substantially higher gas pressures than those used by Gong {\it et al.}\ \cite{gong:2008}  to approach the high-pressure, linear regime. As Fig.~1(b) shows, even at 100~Torr, much of the nonlinear shift is still present.
Several improvements were needed to precisely measure nonlinear shifts of a few Hz on top of background, linear shifts of up to 120~kHz at high pressures.
The most important was the removal of nonlinearity in pressure measurement, which we describe below.
%%%%%%%%%%%%%%%%%%%%%%%
We measured the pressure $p$ with capacitance manometers (MKS Instruments Baratron) to a precision of roughly $\pm 0.002$~Torr and an accuracy of about $\pm0.25\%$.
We found that the true pressure $p$ has a slight quadratic dependence on the pressure $p_{\rm g}$ measured by these gauges.
We estimated the true pressure for our data with the empirical formula,
\begin{align}	\label{pgauge}
p &= r(\alpha) (p_\text{g} + \alpha \, p_\text{g}^2),
\end{align}
where the coefficient $\alpha$ describes the curvature and the coefficient $r = r(\alpha) \approx 1$ minimizes the small linear bias of (\ref{pgauge}) \footnote{We determined $r(\alpha)$ by minimizing the least-squares linear bias $\int (p-p_\text{g})^2 dp_\text{g}$ over a 0--100 Torr gauge range.}.
Using the measured pressures $p_\text{g}$, He, Ne, and N$_2$ appear to have nonlinear shifts, but, using the same empirically determined value of $\alpha$ with (\ref{pgauge}) eliminates these nonlinearities.
We determined the values of $\alpha$ from measurements with He, Ne, and N$_{2}$, for which the effects of vdW molecules, if present at all, appear below our experimental accuracy.
For the Rb gauge, we found $\alpha = (-3.40 \pm 0.22 )\times10^{-5}$~Torr$^{-1}$,
and for the Cs gauge $\alpha = (-1.71 \pm 0.46 )\times10^{-5}$~Torr$^{-1}$.
These values of $\alpha$ appear to have been stable over the course of our measurements.
Additionally, we found that a gauge controller initially used in the Cs system imparted a several-hertz jump near 4 Torr, which disappeared after replacing the controller.

We observed anomalous frequency shifts in some measurements at very low pressures, below about 1 Torr, with gases other than Xe.  These shifts were most noticeable with the Cs system, where they could be as large as $-25$~Hz at 0.7~Torr for He, Ne, or N$_2$.  The shifts were less reproducible in the Rb system, where they were typically less than $5$~Hz at 0.5~Torr in Ar, Kr, Ne, or N$_2$, but were not observed in He.
Except for Rb in Ar, where they were of either sign, these shifts were negative. The origin of these anomalous shifts seems to be a systematic effect, perhaps associated with the very poor signal-to-noise ratios at very low pressures.  Further investigation is needed to identify their cause.

%	FIGURE 3
\begin{figure}[t]
	\centering
	\includegraphics{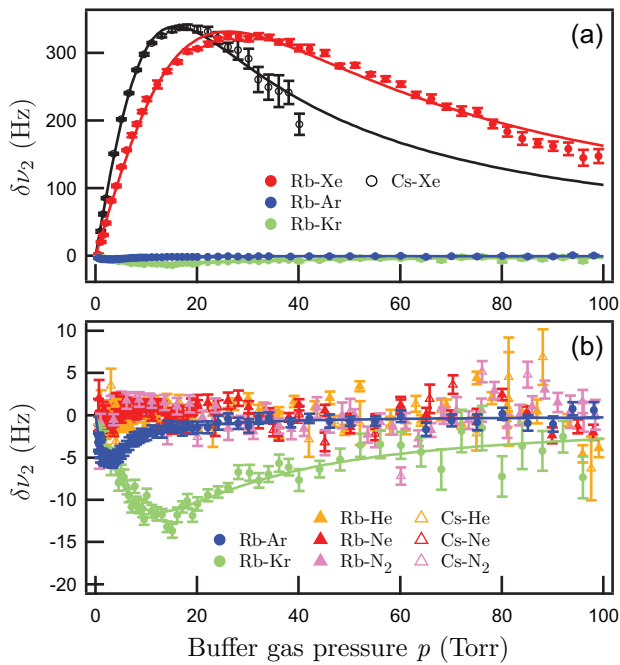}
	\caption{\label{fig:3}(Color online) Summary of the nonlinear shifts $\delta \nu_2$ for \super{87}Rb at 40\degree C and $B=1$ G and for \super{133}Cs at 35\degree C and $B=0.2$ G.  (a) Rb and Cs in Xe, with Rb in Ar and Kr for comparison.  (b) Rb in Ar and Kr, and Rb and Cs in He, Ne, and N\sub{2}. The solid curves are the function $f_0$ of (\ref{f0}) with parameters in Table \ref{tab:1}.}
\end{figure}

Fig.~3 summarizes our experimental measurements.
As Fig.~3(a) shows, the nonlinear shifts in Xe are relatively large and opposite in sign compared to those in Ar and Kr.
Table \ref{tab:1} lists the pressure-independent fit parameters for the data.
The linear shifts $s$ are consistent with previous work \cite{gong:2008, arditicarver:1961, bean:1976}.
For Xe, results are provided for both fit functions (\ref{f0}) and (\ref{f1}).
For Cs in Xe, we fit to $f_1$ of (\ref{f1}) with $r_1$ as a free parameter, since we were unable to estimate $r_1 \phi p$, as we did for Rb, due to a lack of experimental data for $\langle \tau p \rangle$.
Because of this, and the reduced pressure range, the fit returned unrealistic uncertainties, not shown in Table I.  
For Rb in Ar and Kr, we did not fit with $f_1$ of (\ref{f1}) since the effects of the applied field $B$ may be important.
Instead, we fit with $f_0$ of (\ref{f0}) to provide improved results over those of Gong {\it et al.}\ \cite{gong:2008}.
The new values for $Tp^2$ and $\phi p$ agree, with the exception of $\phi p$ for Rb in Kr, which is roughly half the value of $25\pm2.13$~rad~Torr reported by Gong {\it et al.}\ \cite{gong:2008}.  As a result, we suspect that the previous results may be biased by nonlinearity in pressure measurement, which we corrected for in our independent measurements using (\ref{pgauge}).

The formation-rate parameters $Tp^2$ are larger than the corresponding values (in sec Torr$^2$) of $0.016$, $0.0106$, and $0.00429$ for Rb in Ar, Kr, and Xe, respectively, measured by Bouchiat {\it et al.}\ \cite{bouchiat:1972, bouchiat:1975}.  As pointed out by Gong {\it et al.}\ \cite{gong:2008}, this is likely due to a cancellation of contributions from different vibration-rotation states of the vdW molecules.  For Ar and Kr, where the signs of $\phi$ and $s$ are opposite, the shift parameter $\delta A = \delta A(R)$ must change sign as the internuclear separation $R$ increases.  A similar radial dependence for $\delta A$ was measured for K in Ar \cite{kleppner:1976}.  For Xe, however, the signs of $\phi$ and $s$ are the same, which suggests that $\delta A$ may have a qualitatively different shape over the range of $R$ important to vdW molecules.

In summary, we report relatively large nonlinear pressure shifts of $^{87}$Rb and $^{133}$Cs in Xe.  
Discrepancies with the previous model (\ref{f0}) demonstrate the importance of the spin-rotation interaction (\ref{sr}) to nonlinear pressure shifts and to the secondary status of widely used atomic frequency standards.  
We provide an improved model (\ref{f1}) of the shifts for Xe, report improved measurements of the shifts of $^{87}$Rb in Ar and Kr, and report linear shifts for $^{87}$Rb and $^{133}$Cs in Ne as well as in He and N$_2$ to higher pressures than previously measured.  
Further precision measurement of nonlinear shifts should give a better understanding of the detailed physics of vdW molecules.  

%% == Acknowledgments ==
%\begin{acknowledgments}
\vspace{3mm}
The authors are grateful to M.~J.~Souza for making cells, to F.~Gong and S.~W.~Morgan for contributions to the original apparatus, and to J.~H.~Hendricks at NIST and D.~Jacobs at MKS Instruments for helpful discussions about capacitance manometers.  This work was supported by the Air Force Office of Scientific Research, the Department of Defense, and the Department of Energy.
%\end{acknowledgments}

%% == References ==
%\bibliography{references}

\begin{thebibliography}{10}%
\makeatletter
\providecommand \@ifxundefined [1]{%
 \ifx #1\undefined \expandafter \@firstoftwo
 \else \expandafter \@secondoftwo
\fi
}%
\providecommand \@ifnum [1]{%
 \ifnum #1\expandafter \@firstoftwo
 \else \expandafter \@secondoftwo
\fi
}%
\providecommand \enquote [1]{``#1''}%
\providecommand \bibnamefont  [1]{#1}%
\providecommand \bibfnamefont [1]{#1}%
\providecommand \citenamefont [1]{#1}%
\providecommand\href[0]{\@sanitize\@href}%
\providecommand\@href[1]{\endgroup\@@startlink{#1}\endgroup\@@href}%
\providecommand\@@href[1]{#1\@@endlink}%
\providecommand \@sanitize [0]{\begingroup\catcode`\&12\catcode`\#12\relax}%
\@ifxundefined \pdfoutput {\@firstoftwo}{%
 \@ifnum{\z@=\pdfoutput}{\@firstoftwo}{\@secondoftwo}%
}{%
 \providecommand\@@startlink[1]{\leavevmode}%
 \providecommand\@@endlink[0]{}%
}{%
 \providecommand\@@startlink[1]{%
  \leavevmode
  \pdfstartlink
   attr{/Border[0 0 1 ]/H/I/C[0 1 1]}%
   user{/Subtype/Link/A<</Type/Action/S/URI/URI(#1)>>}%
  \relax
 }%
 \providecommand\@@endlink[0]{\pdfendlink}%
}%
\providecommand \url  [0]{\begingroup\@sanitize \@url }%
\providecommand \@url [1]{\endgroup\@href {#1}{\urlprefix}}%
\providecommand \urlprefix [0]{URL }%
\providecommand \Eprint[0]{\href }%
\@ifxundefined \urlstyle {%
  \providecommand \doi [1]{doi:\discretionary{}{}{}#1}%
}{%
  \providecommand \doi [0]{doi:\discretionary{}{}{}\begingroup
  \urlstyle{rm}\Url }%
}%
\providecommand \doibase [0]{http://dx.doi.org/}%
\providecommand \Doi[1]{\href{\doibase#1}}%
\providecommand \bibAnnote [3]{%
  \BibitemShut{#1}%
  \begin{quotation}\noindent
    \textsc{Key:}\ #2\\\textsc{Annotation:}\ #3%
  \end{quotation}%
}%
\providecommand \bibAnnoteFile [2]{%
  \IfFileExists{#2}{\bibAnnote {#1} {#2} {\input{#2}}}{}%
}%
\providecommand \typeout [0]{\immediate \write \m@ne }%
\providecommand \selectlanguage [0]{\@gobble}%
\providecommand \bibinfo [0]{\@secondoftwo}%
\providecommand \bibfield [0]{\@secondoftwo}%
\providecommand \translation [1]{[#1]}%
\providecommand \BibitemOpen[0]{}%
\providecommand \bibitemStop [0]{}%
\providecommand \bibitemNoStop [0]{.\EOS\space}%
\providecommand \EOS [0]{\spacefactor3000\relax}%
\providecommand \BibitemShut [1]{\csname bibitem#1\endcsname}%
%</preamble>
\bibitem{camparo:2007pt}%
  \BibitemOpen
  \bibfield{author}{%
  \bibinfo {author} {\bibfnamefont{J.}~\bibnamefont{Camparo}},\ }%
  \bibfield{journal}{%
  \Doi{10.1063/1.2812121}{\bibinfo {journal} {Physics Today}}\ }%
  \textbf{\bibinfo {volume} {60}},\ \bibinfo {pages} {33} (\bibinfo {year}
  {2007})%
  \bibAnnoteFile{NoStop}{camparo:2007pt}%
\bibitem{vanier:1989}%
  \BibitemOpen
  \bibfield{author}{%
  \bibinfo {author} {\bibfnamefont{J.}~\bibnamefont{Vanier}}\ and\ \bibinfo
  {author} {\bibfnamefont{C.}~\bibnamefont{Audoin}},\ }%
  \emph{\bibinfo {title} {The Quantum Physics of Atomic Frequency Standards}}\
  (\bibinfo {publisher} {Hilger},\ \bibinfo {address} {Philadelphia},\
  \bibinfo {year} {1989})%
  \bibAnnoteFile{NoStop}{vanier:1989}%
\bibitem{walker:1997}%
  \BibitemOpen
  \bibfield{author}{%
  \bibinfo {author} {\bibfnamefont{T.~G.}\ \bibnamefont{Walker}}\ and\ \bibinfo
  {author} {\bibfnamefont{W.}~\bibnamefont{Happer}},\ }%
  \bibfield{journal}{%
  \Doi{10.1103/RevModPhys.69.629}{\bibinfo {journal} {Rev. Mod. Phys.}}\ }%
  \textbf{\bibinfo {volume} {69}},\ \bibinfo {pages} {629} (\bibinfo {year}
  {1997})%
  \bibAnnoteFile{NoStop}{walker:1997}%
\bibitem{Brahms:2010}%
  \BibitemOpen
  \bibfield{author}{%
  \bibinfo {author} {\bibfnamefont{N.}~\bibnamefont{Brahms}}, \bibinfo {author}
  {\bibfnamefont{T.~V.}\ \bibnamefont{Tscherbul}}, \bibinfo {author}
  {\bibfnamefont{P.}~\bibnamefont{Zhang}}, \bibinfo {author}
  {\bibfnamefont{J.}~\bibnamefont{K{\l}os}}, \bibinfo {author}
  {\bibfnamefont{H.~R.}\ \bibnamefont{Sadeghpour}}, \bibinfo {author}
  {\bibfnamefont{A.}~\bibnamefont{Dalgarno}}, \bibinfo {author}
  {\bibfnamefont{J.~M.}\ \bibnamefont{Doyle}},\ and\ \bibinfo {author}
  {\bibfnamefont{T.~G.}\ \bibnamefont{Walker}},\ }%
  \bibfield{journal}{%
  \Doi{10.1103/PhysRevLett.105.033001}{\bibinfo {journal} {Phys. Rev. Lett.}}\
  }%
  \textbf{\bibinfo {volume} {105}},\ \bibinfo {pages} {033001} (\bibinfo {year}
  {2010})%
  \bibAnnoteFile{NoStop}{Brahms:2010}%
\bibitem{gong:2008}%
  \BibitemOpen
  \bibfield{author}{%
  \bibinfo {author} {\bibfnamefont{F.}~\bibnamefont{Gong}}, \bibinfo {author}
  {\bibfnamefont{Y.-Y.}\ \bibnamefont{Jau}},\ and\ \bibinfo {author}
  {\bibfnamefont{W.}~\bibnamefont{Happer}},\ }%
  \bibfield{journal}{%
  \Doi{10.1103/PhysRevLett.100.233002}{\bibinfo {journal} {Phys. Rev. Lett.}}\
  }%
  \textbf{\bibinfo {volume} {100}},\ \bibinfo {pages} {233002} (\bibinfo {year}
  {2008})%
  \bibAnnoteFile{NoStop}{gong:2008}%
\bibitem{pritchard:1976}%
  \BibitemOpen
  \bibfield{author}{%
  \bibinfo {author} {\bibfnamefont{R.}~\bibnamefont{Goldhor}}\ and\ \bibinfo
  {author} {\bibfnamefont{D.}~\bibnamefont{Pritchard}},\ }%
  \bibfield{journal}{%
  \Doi{doi:10.1063/1.432317}{\bibinfo {journal} {J. Chem. Phys.}}\ }%
  \textbf{\bibinfo {volume} {64}},\ \bibinfo {pages} {1189} (\bibinfo {year}
  {1976})%
  \bibAnnoteFile{NoStop}{pritchard:1976}%
\bibitem{wu:1985}%
  \BibitemOpen
  \bibfield{author}{%
  \bibinfo {author} {\bibfnamefont{Z.}~\bibnamefont{Wu}}, \bibinfo {author}
  {\bibfnamefont{T.~G.}\ \bibnamefont{Walker}},\ and\ \bibinfo {author}
  {\bibfnamefont{W.}~\bibnamefont{Happer}},\ }%
  \bibfield{journal}{%
  \Doi{10.1103/PhysRevLett.54.1921}{\bibinfo {journal} {Phys. Rev. Lett.}}\ }%
  \textbf{\bibinfo {volume} {54}},\ \bibinfo {pages} {1921} (\bibinfo {year}
  {1985})%
  \bibAnnoteFile{NoStop}{wu:1985}%
\bibitem{bouchiat:1972}%
  \BibitemOpen
  \bibfield{author}{%
  \bibinfo {author} {\bibfnamefont{M.~A.}\ \bibnamefont{Bouchiat}}, \bibinfo
  {author} {\bibfnamefont{J.}~\bibnamefont{Brossel}},\ and\ \bibinfo {author}
  {\bibfnamefont{L.~C.}\ \bibnamefont{Pottier}},\ }%
  \bibfield{journal}{%
  \Doi{10.1063/1.1677750}{\bibinfo {journal} {J. Chem. Phys.}}\ }%
  \textbf{\bibinfo {volume} {56}},\ \bibinfo {pages} {3703} (\bibinfo {year}
  {1972})%
  \bibAnnoteFile{NoStop}{bouchiat:1972}%
\bibitem{Note1}%
  \BibitemOpen
  \bibinfo {note} {For $^{\protect \text {133}}$Cs, $[W_7,\protect \ldots
  ,W_{-7}]$ = [245, 245, 425, 425, 659, 659, 1231, 1231, 1231, 659, 659, 425,
  425, 245, 245]/9009.}%
  \bibAnnoteFile{Stop}{Note1}%
\bibitem{jau:2004}%
  \BibitemOpen
  \bibfield{author}{%
  \bibinfo {author} {\bibfnamefont{Y.-Y.}\ \bibnamefont{Jau}}, \bibinfo
  {author} {\bibfnamefont{A.~B.}\ \bibnamefont{Post}}, \bibinfo {author}
  {\bibfnamefont{N.~N.}\ \bibnamefont{Kuzma}}, \bibinfo {author}
  {\bibfnamefont{A.~M.}\ \bibnamefont{Braun}}, \bibinfo {author}
  {\bibfnamefont{M.~V.}\ \bibnamefont{Romalis}},\ and\ \bibinfo {author}
  {\bibfnamefont{W.}~\bibnamefont{Happer}},\ }%
  \bibfield{journal}{%
  \Doi{10.1103/PhysRevLett.92.110801}{\bibinfo {journal} {Phys. Rev. Lett.}}\
  }%
  \textbf{\bibinfo {volume} {92}},\ \bibinfo {pages} {110801} (\bibinfo {year}
  {2004})%
  \bibAnnoteFile{NoStop}{jau:2004}%
\bibitem{linparlin:2005}%
  \BibitemOpen
  \bibfield{author}{%
  \bibinfo {author} {\bibfnamefont{A.~V.}\ \bibnamefont{Taichenachev}},
  \bibinfo {author} {\bibfnamefont{V.~I.}\ \bibnamefont{Yudin}}, \bibinfo
  {author} {\bibfnamefont{V.~L.}\ \bibnamefont{Velichansky}},\ and\ \bibinfo
  {author} {\bibfnamefont{S.~A.}\ \bibnamefont{Zibrov}},\ }%
  \bibfield{journal}{%
  \Doi{10.1134/1.2142864}{\bibinfo {journal} {JETP Lett.}}\ }%
  \textbf{\bibinfo {volume} {82}},\ \bibinfo {pages} {398} (\bibinfo {year}
  {2005})%
  \bibAnnoteFile{NoStop}{linparlin:2005}%
\bibitem{Note2}%
  \BibitemOpen
  \bibinfo {note} {We determined $r(\alpha )$ by minimizing the least-squares
  linear bias $\DOTSI \intop \ilimits@ (p-p_\protect \text {g})^2 dp_\protect
  \text {g}$ over a 0--100-Torr gauge range.}%
  \bibAnnoteFile{Stop}{Note2}%
\bibitem{arditicarver:1961}%
  \BibitemOpen
  \bibfield{author}{%
  \bibinfo {author} {\bibfnamefont{M.}~\bibnamefont{Arditi}}\ and\ \bibinfo
  {author} {\bibfnamefont{T.~R.}\ \bibnamefont{Carver}},\ }%
  \bibfield{journal}{%
  \Doi{10.1103/PhysRev.124.800}{\bibinfo {journal} {Phys. Rev.}}\ }%
  \textbf{\bibinfo {volume} {124}},\ \bibinfo {pages} {800} (\bibinfo {year}
  {1961})%
  \bibAnnoteFile{NoStop}{arditicarver:1961}%
\bibitem{bean:1976}%
  \BibitemOpen
  \bibfield{author}{%
  \bibinfo {author} {\bibfnamefont{B.~L.}\ \bibnamefont{Bean}}\ and\ \bibinfo
  {author} {\bibfnamefont{R.~H.}\ \bibnamefont{Lambert}},\ }%
  \bibfield{journal}{%
  \Doi{10.1103/PhysRevA.13.492}{\bibinfo {journal} {Phys. Rev. A}}\ }%
  \textbf{\bibinfo {volume} {13}},\ \bibinfo {pages} {492} (\bibinfo {year}
  {1976})%
  \bibAnnoteFile{NoStop}{bean:1976}%
\bibitem{bouchiat:1975}%
  \BibitemOpen
  \bibfield{author}{%
  \bibinfo {author} {\bibfnamefont{M.~A.}\ \bibnamefont{Bouchiat}}, \bibinfo
  {author} {\bibfnamefont{J.}~\bibnamefont{Brossel}}, \bibinfo {author}
  {\bibfnamefont{P.}~\bibnamefont{Mora}},\ and\ \bibinfo {author}
  {\bibfnamefont{L.}~\bibnamefont{Pottier}},\ }%
  \bibfield{journal}{%
  \Doi{10.1051/jphys:0197500360110107500}{\bibinfo {journal} {J. Phys.
  France}}\ }%
  \textbf{\bibinfo {volume} {36}},\ \bibinfo {pages} {1075} (\bibinfo {year}
  {1975})%
  \bibAnnoteFile{NoStop}{bouchiat:1975}%
\bibitem{kleppner:1976}%
  \BibitemOpen
  \bibfield{author}{%
  \bibinfo {author} {\bibfnamefont{R.~R.}\ \bibnamefont{Freeman}}, \bibinfo
  {author} {\bibfnamefont{D.~E.}\ \bibnamefont{Pritchard}},\ and\ \bibinfo
  {author} {\bibfnamefont{D.}~\bibnamefont{Kleppner}},\ }%
  \bibfield{journal}{%
  \Doi{10.1103/PhysRevA.13.907}{\bibinfo {journal} {Phys. Rev. A}}\ }%
  \textbf{\bibinfo {volume} {13}},\ \bibinfo {pages} {907} (\bibinfo {year}
  {1976})%
  \bibAnnoteFile{NoStop}{kleppner:1976}%
\end{thebibliography}

%DON'T FORGET TO PASTE bbl FILE CONTENTS HERE!

%Merlin.mbs v4.21 2009-07-09.
%

%% == Figure Captions ==
%\newpage
%
% Use the figure* environment if the figure should span across the
% entire page. There is no need to do explicit centering.

\end{document}